\documentclass{aa}
\usepackage{amssymb,amsmath}
\usepackage{graphicx,epstopdf}
\usepackage{natbib}
\usepackage{bm}
\usepackage{ulem}
\bibpunct{(}{)}{;}{a}{}{,} 
\usepackage{color}
\def\lesssim{\;\raise0.3ex\hbox{$<$\kern-0.75em\raise-1.1ex\hbox{$\sim$}}\;}
\def\gtrsim{\;\raise0.3ex\hbox{$>$\kern-0.75em\raise-1.1ex\hbox{$\sim$}}\;}

\def\msun{M_\odot}

\def\gcm{g~cm$^{-3}$}
\def\beq{\begin{equation}}
\def\eeq{\end{equation}}
\begin{document}
\title{Accreting neutron stars from the nuclear energy-density functional theory. II. Equation of state and global properties
\thanks{{The tables of the equations of state are available at the CDS via anonymous ftp to \url{cdsarc.u-strasbg.fr} (130.79.128.5) or via
\url{http://cdsweb.u-strasbg.fr/cgi-bin/qcat?J/A+A/}}}
}

\author{A.~F.  Fantina\inst{1,2},
J.~L.  Zdunik\inst{3}, 
N.  Chamel\inst{2},
J.~M.  Pearson\inst{4},
L.  Suleiman\inst{3,5},
S.  Goriely\inst{2}
}

\institute{Grand Acc\'el\'erateur National d'Ions Lourds (GANIL), CEA/DRF - CNRS/IN2P3, Boulevard Henri Becquerel, 14076 Caen, France
\email{anthea.fantina@ganil.fr}
\and Institut d'Astronomie et d'Astrophysique, CP-226, Universit\'e Libre de Bruxelles, 1050 Brussels, Belgium 
\and N. Copernicus Astronomical Center, Polish Academy of Sciences, Bartycka 18, PL-00-716 Warszawa, Poland 
\and D\'ept. de Physique, Universit\'e de Montr\'eal, Montr\'eal (Qu\'ebec), H3C 3J7 Canada \and Laboratoire Univers et Th\'eories, Observatoire de Paris, Universit\'e PSL, Universit\'e Paris Cit\'e, CNRS, F-92190 Meudon, France
}

\date{Received xxx Accepted xxx}

\abstract{The accretion of matter onto the surface of a neutron star in a low-mass X-ray binary triggers X-ray bursts, whose ashes are buried and further processed thus altering the composition and the properties of the stellar crust. 
}
{In this second paper of a series, the impact of accretion on the equation of state and on the global properties of neutron stars is studied  in the framework of the nuclear energy-density functional theory.
}
{Considering ashes made of $^{56}$Fe, we calculated the equations of state using the same Brussels-Montreal nuclear energy-density functionals BSk19, BSk20, and BSk21, as those already employed for determining the crustal heating in our previous study for the same ashes. All regions of accreting neutron stars were treated in a unified and thermodynamically consistent way. With these equations of state, we determined the mass, radius, moment of inertia, and tidal deformability of accreted neutron stars and compared with catalyzed neutron stars for which unified equations of state based on the same functionals are available. 
}
{The equation of state of accreted neutron stars is found to be significantly stiffer than that of catalyzed matter, with an adiabatic index $\Gamma \approx 4/3$ throughout the crust. For this reason, accreting neutron stars have larger radii. However, their crustal moment of inertia and their tidal deformability are hardly changed provided density discontinuities at the interface between adjacent crustal layers are properly taken into account. 
}
{The enhancement of the stiffness of the equation of state of accreting neutron stars is mainly a consequence of nuclear shell effects, thus confirming the importance of a quantum treatment as stressed in our first study. With our previous calculations of crustal heating using the same functionals, we have thus obtained consistent microscopic inputs for simulations of accreting neutron stars. 
}


\titlerunning{EoS of accreting neutron stars}
\authorrunning{Fantina et al.}

\maketitle

\section{Introduction}
\label{sect:introd}

The accretion of matter onto a neutron star (NS) in a low-mass X-ray binary induces thermonuclear explosions with total energies $10^{39}-10^{40}$~erg observed as X-ray bursts lasting a few tens of seconds and with a typical recurrence time of hours to days~(see e.g. \cite{strohmayer2006} for a review and \cite{galloway2020} for a recent compilation of observations). The unstable carbon burning in deeper layers is thought to power less frequent superbursts with energies $\sim 10^{42}$~erg. On a longer timescale, the ashes from X-ray bursts may be further reprocessed by electron captures, neutron emissions, and neutron captures, and possibly pycnonuclear fusion reactions as they slowly sink inside the NS (see, e.g., \cite{meisel2018} for a recent review). Depending on the duration of accretion episodes, the original crust of the NS can be partially or fully replaced~(\citet{tauris2012}; see also the discussion in \cite{fantina2018}) and this may radically change its properties, as first shown by \cite{hz1990a}. 

Simulations of accreting NSs usually rely on the equation of state (EoS) of \cite{hz1990b} for the accreting crust together with the heat sources obtained by \cite{hz1990a} (or the more recent calculations of \cite{HZ2003,HZ2008}). These results, which are based on the liquid-drop model, are then combined with some EoS for the deepest layers of the inner crust and for the core. However, significant errors on the global structure of the NS and on its tidal deformability may arise from the use of thermodynamically inconsistent EoSs~\citep{fortin2016,ferreira2020,suleiman2021}.

In our previous study, we calculated the heat released in accreting NS crusts and the location of the heat sources  using accurately calibrated Brussels-Montreal nuclear energy-density functionals (EDFs) taking into account nuclear shell effects perturbatively~\citep{fantina2018}. We found that the composition of the crust of an accreted NS can differ substantially from that of an isolated NS made of cold catalyzed matter. In this second paper, we present the corresponding EoSs. Both the outer and inner regions of accreted NS crusts are treated consistently within the same microscopic framework using the same EDFs as in \cite{fantina2018}. Together with the EoSs of the liquid core previously calculated in \cite{goriely2010}, unified EoSs describing all regions of accreted NSs with the same EDFs are constructed. 

After briefly recalling the main features of the microscopic model of accreted NS crusts and of the adopted EDFs in Sect.~\ref{sec:micro}, the calculated EoSs are presented in Sect.~\ref{sec:EOS}. The global properties of accreted NSs are discussed in Sect.~\ref{sec:ns-prop} and conclusions are given in Sect.~\ref{sect:discussion}.

\section{Model of accreted neutron-star crusts}
\label{sec:micro}

\subsection{Main assumptions}
\label{sec:assumpt}

We recall here the main assumptions on how we model the crust of an accreting NS, by which we mean the region of the NS below the envelope and whose density is higher than $10^6$~\gcm. 
Because of the relatively low temperatures generally prevailing in accreting NS crusts, namely $T < 10^9$~K (see e.g. \cite{lrr}), the thermal contributions to the thermodynamic potentials can be neglected.
For such temperatures, except for the shallowest regions, the innermost layers are expected to be in the solid state \citep{fantina2020, carreau2020}. The composition and structure of each layer were determined in the one-component plasma approximation: at each pressure $P$, matter is fully described by the proton number $Z$ and the total nucleon number $A_{\rm cell}$ in a single Wigner-Seitz (WS) cell. In the outer crust, $A_{\rm cell}$ reduces to the mass number $A$ of the corresponding nuclei, while in the inner crust $A_{\rm cell}$ also accounts for free neutrons. 
Unlike cold catalyzed matter, accreting NS crusts are not in full thermodynamic equilibrium and reactions can thus occur. 
The compression of the ashes from X-ray bursts  may trigger a series of electron captures, by which the nucleus $(A,Z)$ transforms into a nucleus $(A,Z-1)$ with the emission of an electron neutrino. 
The one-component plasma approach induces discontinuities in the neutron density and the neutron chemical potential, while in fact the onset of neutron emission occurs at lower density and pressure than those predicted by the mean-nucleus approximation (see \cite{chamel2015} for details). However, we did not consider neutron diffusion \citep{shch2018, chu2020,gus2020}, for consistency with our previous calculations of the heating~\citep{fantina2018}. 
With increasing density and decreasing proton number, neighbouring nuclei may overcome their Coulomb barrier and fuse. 
However, large uncertainties reside in the determination of the rates of these pycnonuclear reactions \citep[see e.g.][]{yak06b}.
We thus simply assumed that these processes occur whenever $Z$ reaches the minimum value $Z_\textrm{min}=8$ (see also the discussion in Sect.~3.2 in \citet{fantina2018}). As in \cite{fantina2018}, we supposed that the crust is fully accreted.

\subsection{Brussels-Montreal nuclear energy-density functionals}
\label{sec:EDF}

Our models of accreted NS crusts are based on the nuclear EDF theory~\citep{bhr03,sto07}, which has proved successful in describing finite nuclei, such as those encountered in the outer crust of a NS \citep{chamel2020}, but also provides a consistent treatment of the inner crust, where neutron-proton clusters coexist with a neutron liquid. 
In this quantum mechanical approach, bound and unbound neutrons are consistently calculated.
They do not need to be considered separately as in the liquid-drop treatment of \cite{hz1990a,hz1990b,HZ2003,HZ2008}.
Moreover, the EDF approach is suitable to describe homogeneous matter, thus allowing for a unified description of all regions of a NS. 
Here, we employ the same Brussels-Montreal EDFs labelled BSk19, BSk20, and BSk21 \citep{goriely2010} as in \citet{fantina2018}.

The Brussels-Montreal EDFs BSk19, BSk20, and BSk21 are based on generalised Skyrme effective nucleon-nucleon interactions \citep{chamel2009, goriely2010}, supplemented with a microscopic contact pairing interaction \citep{chamel2010}. 
These EDFs were fitted to the 2149 measured masses of nuclei with neutron and proton numbers, $N,Z \geq 8$, given in the 2003 Atomic Mass Evaluation (AME) \citep{audi2003}, with a root-mean square (rms) deviation as low as 0.58~MeV for the three functionals, while ensuring at the same time an optimal fit to charge radii. 
Incidentally, these models also yield an equally good fit to the 
experimental masses of nuclei with $N,Z \geq 8$ given in the latest AME~\citep{huang2021,wang2021}. 
The masses of bound nuclei were obtained by adding to the Hartree-Fock-Bogoliubov (HFB) energy a phenomenological Wigner term and correction term for the collective energy. 

The BSk19, BSk20, and BSk21 EDFs were simultaneously constrained to reproduce various properties of homogeneous nuclear matter as obtained from many-body calculations using realistic two- and three- nucleon interactions. 
Specifically, these EDFs were fitted to three different neutron-matter EoSs, reflecting the current uncertainties of the high-density behaviour of dense matter. 
BSk19 was adjusted to the `soft' EoS of neutron matter of \citet{fp1981} obtained from the realistic Urbana v14 nucleon-nucleon force with the three-body force TNI, BSk20 was fitted to the EoS of \citet{apr1998} labelled ``A18 + $\delta v$ + UIX'', whereas BSk21 was constrained to reproduce the `stiff' EoS labelled ``V18'' from \citet{lischulze2008}. 
All three neutron-matter EoSs are consistent at the low densities prevailing in nuclei and in NS crusts and are compatible with the more recent calculations based on auxiliary field diffusion Monte Carlo method and chiral effective field theory (see \citet{fantina2014} and Fig.~13 in \citet{fantina2018}).

Furthermore: 
(i) the symmetry energy coefficient was set to $J$ = 30 MeV for the three functionals \citep[see e.g.][for a discussion of experimental and theoretical estimates of $J$]{goriely2010, tsang2012, baldo2016, tews2017, burgio2018},
(ii) the incompressibility $K_v$ of symmetric nuclear matter at saturation was required to fall in the experimental range $240\pm10$~MeV~\citep{col04,sto14}, 
(iii) the ratio of the isoscalar effective mass to bare nucleon mass in symmetric nuclear matter at saturation was set to the realistic value of 0.8 \citep{goriely2003}, 
(iv) the isovector effective mass was found to be smaller than the isoscalar effective mass, in agreement with both experiments and many-body calculations \citep[see][for a discussion]{goriely2010}, 
(v) a qualitatively realistic distribution of the potential energy among the four spin-isospin channels in nuclear matter was obtained, and 
(vi) spurious spin and spin-isospin instabilities in nuclear matter that generally plagued earlier Skyrme functionals were eliminated for all densities prevailing in NSs \citep{chamel2009, chamelgoriely2010}. Finally, all these functionals are consistent with constraints on the EoS of symmetric nuclear matter inferred from heavy-ion collisions \citep{daniel2002, lynch2009}.

The BSk19-21 EDFs have been already applied to calculate the structure, the composition, and the EoS of non-accreted NSs under the cold catalyzed matter hypothesis \citep{pearson2011,pearson2012} and analytical representations of these EoSs have also been obtained by \citet{potekhin2013}. 
Even though the EoS based on the BSk19 EDF fails to explain the existence of the most massive observed NSs \citep{chamel2011} (see also \citet{fantina2013} for a discussion of various astrophysical constraints), such a soft EoS seems to be favoured by the analyses of K$^+$ production \citep{fuchs2001, sturm2001, hartnack2006} and $\pi^-/\pi^+$ production ratio \citep{xiao2009} in heavy-ion collisions. 
This EoS also appears to be supported by the LIGO-Virgo data from the gravitational-wave signal GW170817 emitted by the merger of intermediate mass NSs \citep{perot2019}.
We have thus employed these three EDFs to calculate the EoS of all regions of accreted NSs, as described in the following, and we have studied their impact on NS properties.

\subsection{Outer crust}
\label{sec:outcrust}

We assumed that the outer crust is made of fully ionised atoms embedded in a charge-neutralising degenerate electron background.
Since in hydrostatic equilibrium the pressure $P$ varies continuously throughout the star, the suitable thermodynamic potential for determining the equilibrium structure of the crust is the Gibbs free energy per nucleon $g(A,Z,P)$. 
The composition and EoS in each layer of pressure $P$ were thus found by minimising $g$, given by
\beq
g = \frac{ \mathcal{E} + P}{n} \ ,
\label{eq:g}
\eeq 
$\mathcal{E}=\rho c^2$ being the internal energy density and $n$ the average baryon number density, keeping the mass number $A$ fixed (recalling that in the outer crust the number of nucleons in the cell is equal to the nucleus mass number, i.e. $A_{\rm cell} = A$). 

For the specific expressions of $\mathcal{E}$ and $P$, we followed \citet{pearson2011}.
To recall the main points, the internal energy per nucleon $\mathcal{E}/n$ in the WS cell is given by
\beq
\frac{\mathcal{E}}{n} = \frac{M^\prime(A,Z) c^2}{A} +  \frac{\mathcal{E}_L}{n} + \frac{\mathcal{E}_e(n_e)}{n}  \ ,
\label{eq:e}
\eeq
where $M^\prime(A,Z)$ is the nuclear mass, $c$ is the speed of light, and $\mathcal{E}_L$ is the lattice energy density,
\beq
\mathcal{E}_L =  C e^2 Z^{2/3} n_e^{4/3} \ ,
\label{eq:EL}
\eeq
with $e$ the elementary charge, and we adopted the WS estimate $C=-1.4508$ for the lattice constant \citep{sal1961}. 
The last term in Eq.~(\ref{eq:e}), $\mathcal{E}_e$, is the energy density of electrons, which accounts for the energy density of a free uniform electron gas of number density $n_e$ (see \citet{coxgiuli} for complete expressions), including the exchange and polarisation contributions, for which we used Eqs.~(1), (6), and (8) in \cite{chamel2016b}. 
We therefore neglected the electron correlation correction, as well as the finite-size and zero-point contributions, which were shown to be negligible \citep{pearson2011}. 
The electron number density is related to the baryon number density by $n_e=(Z/A)n$ to ensure electric charge neutrality.

Similarly, for the pressure we have
\beq
P = P_e + P_L  \ ,
\label{eq:P}
\eeq
with $P_e$ the pressure of the electron (including exchange and charge-polarisation corrections) and $P_L = \mathcal{E}_L/3$ the lattice pressure, the nuclei exerting no pressure in cold matter.

The only microscopic inputs of this model are the nuclear masses $M^\prime(A,Z)$, that were obtained from the atomic masses $M(A,Z)$ after subtracting out the binding energy of atomic electrons (see Eq.~(A4) in \cite{lpt03}). 
For consistency with our previous study~\citep{fantina2018}, we made use of the experimental atomic masses from the 2016 AME \citep{ame16}, whenever available\footnote{Only measured atomic masses were used. Recommended values were not considered.}, complemented with the Brussels-Montreal HFB atomic masses from the corresponding functional, also available from the BRUSLIB database \citep{bruslib}.

With increasing pressure, $Z$ remains unchanged until $P$ reaches the threshold pressure $P_\beta$ for the onset of electron captures and such that $g(A,Z,P_\beta)=g(A,Z-1,P_\beta)$. 
Accurate analytical formulae for $P_\beta$ and the corresponding density $\rho_\beta$ were obtained by \cite{chamel2015b,chamel2016b}. The daughter nucleus $(A,Z-1)$ is generally unstable and further electron captures occur until the proton number reaches some value $Z_0<Z$ such that  $g(A,Z_0,P_\beta)<g(A,Z_0-1,P_\beta)$, which corresponds to a local minimum of the Gibbs free energy per nucleon at pressure $P_\beta$. 

The onset of neutron drip was determined as in \citet{chamel2015}, thus ensuring the thermodynamic consistency across the boundary between the outer and inner crusts: the proton number of the dripping nucleus is given by the highest value of $Z$ (below that of the initial ashes) for which the $\Delta N$ neutron separation energy is negative, i.e.
\beq
M^\prime(A - \Delta N,Z-1) < M^\prime(A,Z-1) + \Delta N m_n \ ,
\eeq
$m_n$ being the neutron mass.
The properties of the accreted crust at the neutron-drip transition are summarised in Table~\ref{tab:ndrip}, where we list, for the considered models, the proton number $Z$ of the dripping nucleus, the number $\Delta N$ of emitted neutrons, the mass-energy density $\rho_{\rm drip}$, and the corresponding pressure $P_{\rm drip}$.

\begin{table}
\centering
\caption{Neutron-drip transition in the crust of fully accreted NSs, as predicted by HFB-19, HFB-20, and HFB-21 nuclear mass models for $^{56}$Fe ashes: atomic number $Z$ of the dripping nucleus, number $\Delta N$ of emitted neutrons, mass-energy density $\rho_{\rm drip}$ and corresponding pressure $P_{\rm drip}$. }\smallskip
\label{tab:ndrip}
\begin{tabular}{ccccc}
\hline \hline \noalign {\smallskip}
&$Z$ &$\Delta N$ &$\rho_{\rm drip}$  & $P_{\rm drip}$ \\
&&&{\tiny ($10^{11}$ g\,cm$^{-3}$)} & {\tiny ($10^{29}$ dyn\,cm$^{-2}$)}\\
\hline \noalign {\smallskip}
 HFB-19 & 18 & 1 &4.48 & 9.02 \\
 HFB-20 & 18 & 1 & 4.50 & 9.06 \\
 HFB-21 & 18 & 1 & 4.38 & 8.75 \\
 \hline
\end{tabular}
\end{table}

\subsection{Inner crust}
\label{sec:inncrust}

For the inner crust, we calculated the EoS using the 4th order Extended Thomas-Fermi (ETF) approach, with proton shell corrections added perturbatively via the Strutinsky integral (SI) theorem  (see \citet{onsi2008} and \citet{pearson2012} for details). Neutron shell corrections, which are expected to be much smaller than proton shell corrections~\citep{pearson2022}, were neglected. This so-called ETFSI approach is a computationally high-speed approximation of the fully self-consistent HFB method. 
We employed the same EDFs as those underlying the HFB nuclear mass models used for the outer crust, thus ensuring a consistent description of both the outer and the inner regions of the crust.  

We further assumed that nuclear clusters remain spherical and we calculated the Coulomb energy in the WS approximation. Indeed, the presence of so-called nuclear `pasta' has recently been found to be marginal within the ETFSI approach~\citep{pearson2022}. 
In order to further reduce the computation time, the nucleon density distributions in the WS cell were parametrised as 
$n_q (r) = n_{B q} + n_{\Lambda q} f_q (r)$,
where $q = n,p$ for neutrons and protons respectively, $r$ is the radial distance from the centre of the spherical cell, $n_{B q}$ is the background density, and the second term accounts for the presence of inhomogeneities, with $n_{\Lambda q}$ a constant term and $f_q(r)$ a damped Fermi function ensuring that density derivatives vanish at the surface of the cell, thus providing a smooth matching of the nucleon distributions between adjacent cells~\citep{onsi2008}. 

Before the onset of pycnonuclear reactions, $A_{\rm cell}$ remains unchanged with increasing pressure whereas $Z$ can decrease if electron captures occur. 
The EoS can in principle be determined by minimising the baryon chemical potential, which is equal to the Gibbs free energy per nucleon $g$, with respect to all the parameters of the WS cell keeping $A_{\rm cell}$ fixed.  
We note that, because of the presence of the free neutron gas, $A_{\rm cell} > A$ with the number $A$ of nucleons in clusters defined by
\begin{equation}
A =Z + 4\pi\,n_{\Lambda n}\int_0^{R_{\rm{cell}}} r^2f_n(r)dr \,   ,
\end{equation}
with $R_{\rm cell}$ the WS cell radius.
However, this procedure would be computationally very costly. 
Instead, we proceeded as follows: 
(1) We first minimised the internal energy $\mathcal{E}/n$ per nucleon at constant average baryon number density $n$, for given values of the proton number $Z$ and nucleon number $A_\textrm{cell}$ \citep[see][for details]{pearson2012};
(2) For this set of parameters $(n,Z,A_{\rm cell})$, we calculated the corresponding pressure $P$ and the Gibbs free energy per nucleon $g$; 
(3) We repeated the calculations for different values of $n$ and $Z$, thus allowing us to construct $g$ as a function of $Z$ and $P$ for the given (fixed) value of $A_\textrm{cell}$.

\section{EoS for accreted neutron-star crusts}
\label{sec:EOS}

We calculated the composition and the EoSs of fully accreted NS crusts for the EDFs BSk19, BSk20, and BSk21, considering X-ray burst ashes made of $^{56}$Fe as in our previous study~\citep{fantina2018}.

\begin{figure*}
\resizebox{0.49\hsize}{!}{\includegraphics{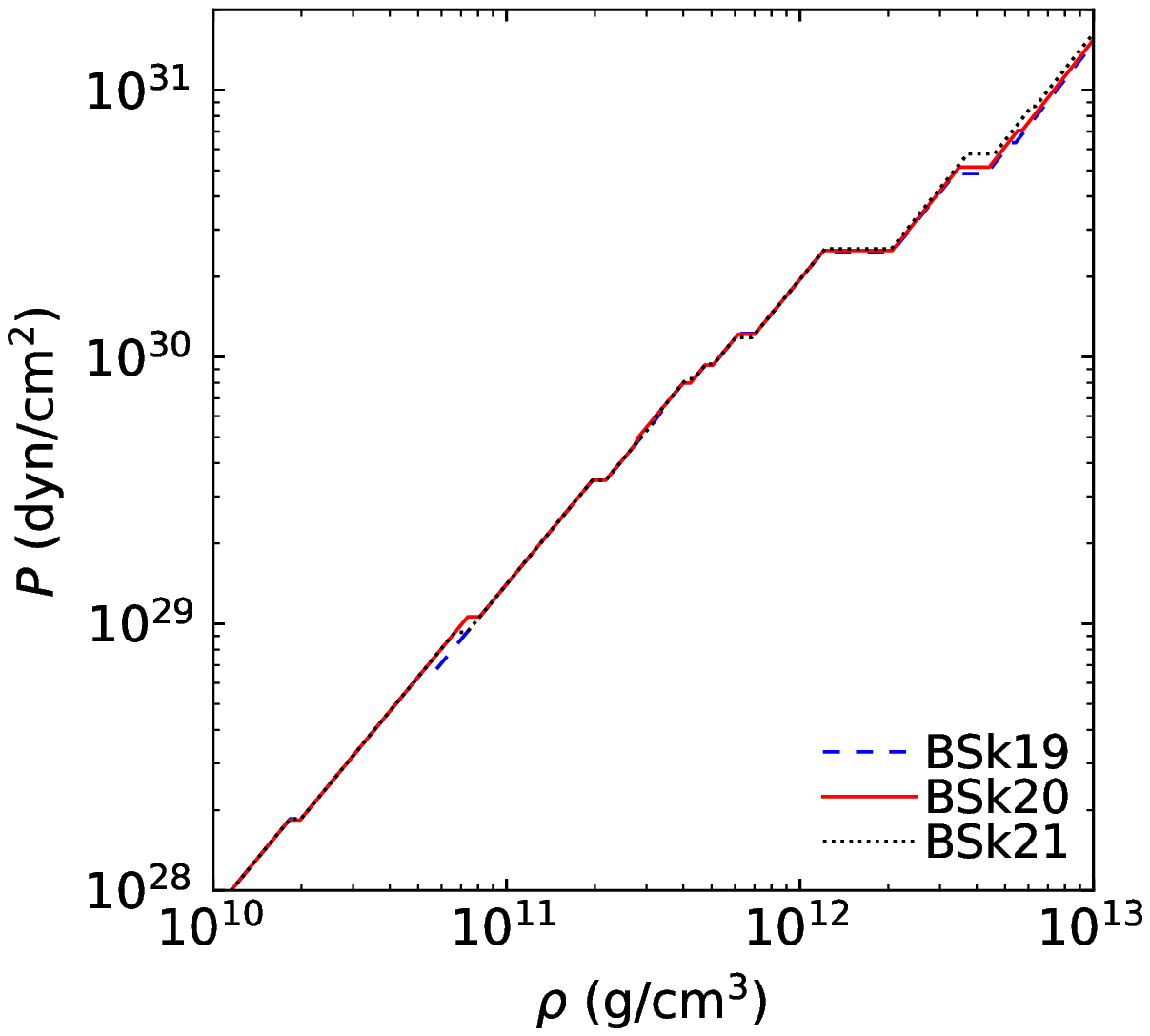}}
\resizebox{0.48\hsize}{!}{\includegraphics{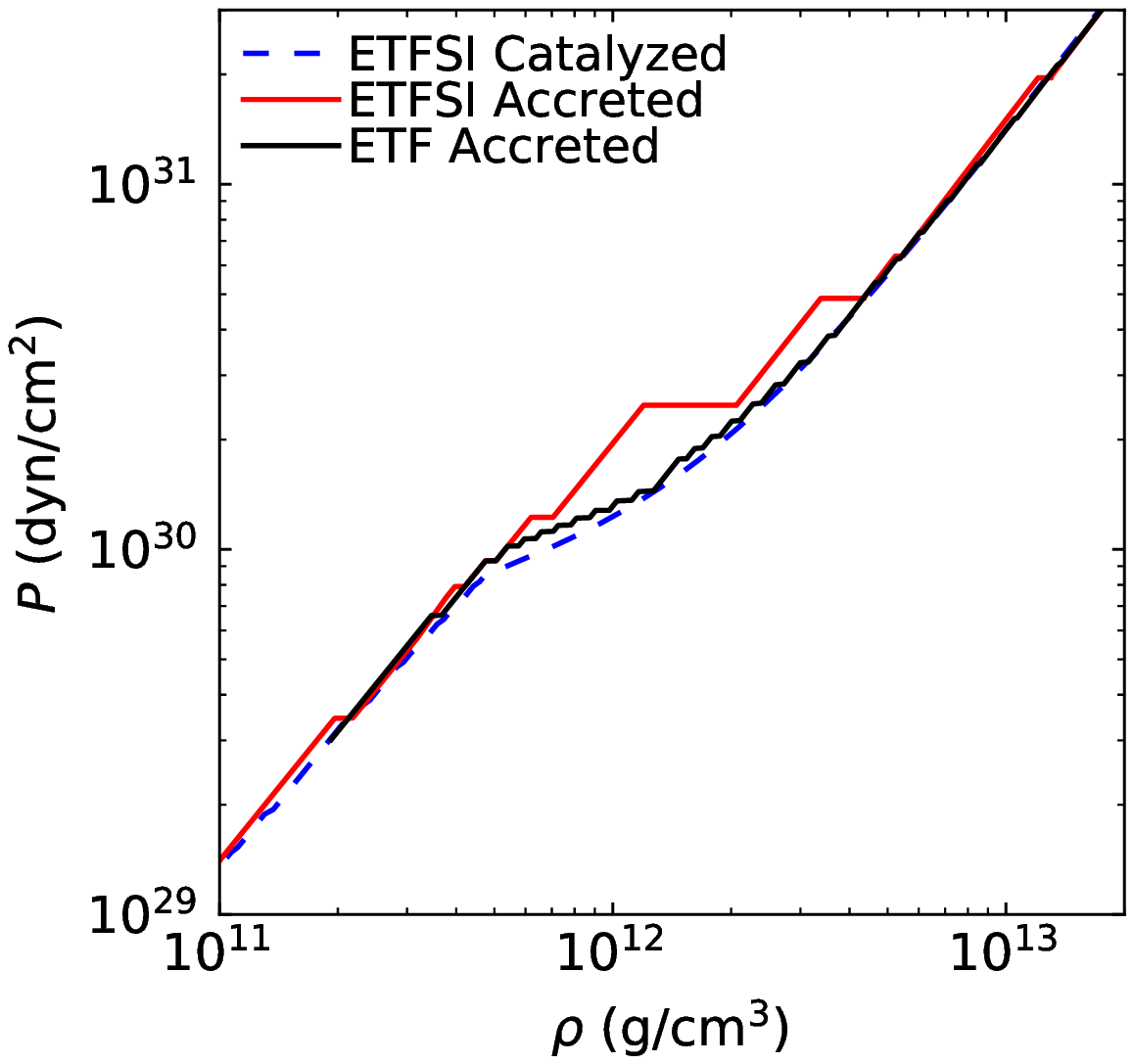}} 
\caption{Pressure versus mass-energy density (in cgs units) in the NS crust. 
Left panel: predictions from the models based on the BSk19, BSk20, and BSk21 EDFs in accreted NS crust. 
Right panel: predictions from the models based on the BSk19 EDF comparing two different treatments for the inner crust: ETFSI and ETF. For comparison, results for the catalyzed crust are also shown. 
See text for details.}
\label{Fig:EOS_192021+ETFsep}
\end{figure*}

\subsection{Pressure-density relation}

Results are plotted in Fig.~\ref{Fig:EOS_192021+ETFsep}. 
The EoS tables containing the number density $n$, the mass-energy density $\rho$, and the pressure $P$  for the three EDFs are available at the CDS.
As we can see from the left panel of Fig.~\ref{Fig:EOS_192021+ETFsep}, the three EoSs exhibit a similar behaviour, with steps associated to changes in the composition induced by electron captures and/or pycnonuclear reactions \citep{fantina2018}. 
On the other hand, remarkable differences are observed between the EoS of accreted and catalyzed crusts, the former being notably stiffer than the latter, particularly in the density region between the neutron drip and $ \approx 10^{13}$~g~cm$^{-3}$. 
This is clearly seen from the right panel of Fig.~\ref{Fig:EOS_192021+ETFsep}, where the EoS based on the BSk19 EDF is shown for both catalyzed (blue dashed line) and accreted (red solid line) crusts, as an illustrative example (the other EDFs yielding qualitatively similar results).
One can also notice that, at higher densities, $\rho \gtrsim 10^{13}$~g~cm$^{-3}$ (corresponding to $P \gtrsim 5 \times 10^{31}\,{\rm dyn\ cm^{-2}}$), the two EoSs merge. 
The global properties of accreted NSs can thus be fairly accurately calculated by combining the EoS of accreted crust with that of catalyzed matter for densities beyond $\rho \gtrsim 10^{13}$~g~cm$^{-3}$ (deep inner crust and core, see Sect.~\ref{sec:ns-prop}). 

These calculations assumed that the crust is fully accreted and stationary.
This assumption is valid if the star has accreted for a long enough time, approximately given by \citep{suleiman2022}
\begin{equation}
\tau_{\rm acc}\simeq 4.8 \frac{P_{31}}{\dot M_{-10}} \frac{R_6^4}{M/\msun}\sqrt{1-\frac{2GM}{Rc^2}}\,{\rm Myr} \ ,
\end{equation}
where $P_{31}$ is the pressure in $10^{31}\,{\rm dyn\ cm^{-2}}$ at the crust-core transition, $\dot M_{-10}$ is the accretion rate in $10^{-10}\,\msun/{\rm yr}$ ($\msun$ being the mass of our Sun), and $R_6=R/(10\,{\rm km})$. 
Setting $P_{31} = 60$ \citep{pearson2012}, $M = 1.4\ \msun$, and $R=12$~km, we find $\tau_{\rm acc} \sim 430\,{\rm Myr}/\dot M_{-10}$.
On longer timescales, the EoS remains unchanged with further accretion and is therefore well defined.
On the other hand, on a shorter accretion timescale, the NS crust is only partially replaced by the accreted material and a different treatment is required \citep{suleiman2022}.

To assess the importance of shell effects on the EoS, we also display the results obtained using the ETF approach (black solid line in the right panel of Fig.~\ref{Fig:EOS_192021+ETFsep}).
As previously discussed in \citet{fantina2018}, the compressed matter element undergoes in this case successive electron captures until the proton number is low enough to allow for pycnonuclear reactions to occur.
As a consequence, variations of $Z$ are more continuous and jumps in density are much smaller with respect to those obtained with the full ETFSI treatment. 
Even though the composition was found to be radically different, we showed that the proton fractions $Z/A_{\rm cell}$ are comparable (see Fig.~4 in \citet{fantina2018}).
Since the pressure is mainly determined by electrons and $\rho \approx n_e m_u A_{\rm cell}/Z$, with $m_u$ the unified atomic mass unit (recalling that $n_e = n Z/A_{\rm cell} $ due to electric charge neutrality), the EoS of accreted NS crusts is thus expected to be very similar to that of catalyzed crusts when shell corrections are neglected, as can be seen in Fig.~\ref{Fig:EOS_192021+ETFsep}. 
This shows that nuclear shell effects are important not only for determining the composition but also for the EoS of accreted crusts.

\subsection{Adiabatic index}
\label{sect:adindex}

\begin{figure}
\begin{center}
\includegraphics[width=.5\textwidth,angle=0]{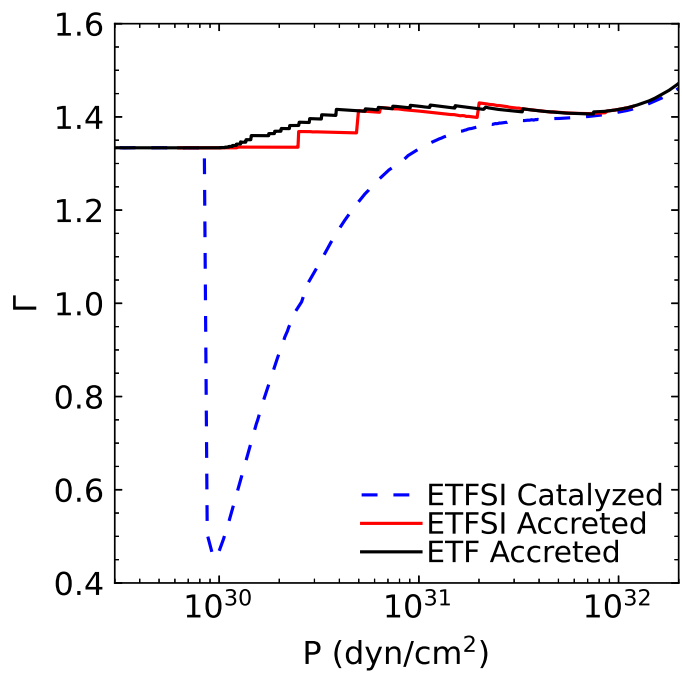}
\end{center}
\caption{Adiabatic index $\Gamma= {\rm d} \log(P)/ {\rm d}\log(n)$ versus pressure (in cgs units) for the EoS based on the BSk19 EDF, presented in the right panel of Fig. \ref{Fig:EOS_192021+ETFsep}. 
}
\label{fig:adindex}
\end{figure}

To better see the differences between the EoS of accreted and catalyzed crusts, we calculated the adiabatic index, 
\begin{equation}
 \Gamma=\frac{{\rm d} \log(P)}{{\rm d} \log(n)} = \frac{n}{P} \frac{ {\rm d} P}{ {\rm d} n} = \frac{\rho + P/c^2}{P} \frac{ {\rm d} P}{ {\rm d} \rho} \ .
\label{eq:gameos}
\end{equation}
Results are plotted in Fig.~\ref{fig:adindex} for models based on the BSk19 EDF. 
We focus on pressures around $P \approx 10^{30}-10^{31}$~dyn~cm$^{-2}$ (or, equivalently, $\rho \approx 10^{12}-10^{13}$~g~cm$^{-3}$) corresponding to the largest deviations in Fig.~\ref{Fig:EOS_192021+ETFsep}. 

Assuming that the pressure in the shallow layers of the inner crust remains mainly determined by electrons, we have $P\approx P(n_e) = P(n Z/A_{\rm cell})$. The adiabatic index can thus be expressed as 
\begin{eqnarray}
\label{eq:approx-gameos}
\Gamma&\approx &\frac{\partial \log(P)}{\partial \log(n)}\biggr\vert_{Z/A_{\rm cell}} +\frac{\partial \log(P)}{\partial \log(Z/A_{\rm cell})}\biggr\vert_{n} \frac{ {\rm d} \log(Z/A_{\rm cell})}{ {\rm d} \log(n)} \nonumber \\
&\approx & \frac{4}{3}\biggl[1+\frac{ {\rm d} \log(Z/A_{\rm cell})}{ {\rm d} \log(n)}\biggr]\, , 
\end{eqnarray}
where in the last equality we have considered the pressure of an ultrarelativistic electron Fermi gas
\begin{equation} 
P\approx \frac{\hbar c (3\pi^2)^{1/3}}{4}\left(\frac{Z}{A_{\rm cell}} n \right)^{4/3} \, ,
\end{equation}
$\hbar$ being the Dirac–Planck constant.
In catalyzed crusts, $Z/A_{\rm cell}$ decreases continuously with increasing $n$ so that ${ {\rm d} (Z/A_{\rm cell})}/{ {\rm d} \log(n)}<0$, see e.g. Fig.~4 in \cite{fantina2018}. After the onset of neutron drip, $Z/A_{\rm cell}$ falls rapidly leading to a sharp drop of $\Gamma$, as shown by the blue dashed line in Fig.~\ref{fig:adindex}. In accreted crusts, both $Z$ and $A_{\rm cell}$ remain unchanged until electrons are captured by clusters or pycnonuclear fusion occurs. Therefore, the second term in Eq.~\eqref{eq:approx-gameos} vanishes beyond the neutron-drip point and $\Gamma$ remains equal to $4/3$ as in the densest layers of the outer crust. For the same reason, the adiabatic indices obtained within the ETF (black solid line) and ETFSI (red solid line) approaches are comparable. With further increase of the density, the neutron contribution to the pressure can no longer be ignored and $\Gamma\gtrsim 4/3$. 
This shows that the polytropic approximation according to which $\Gamma$ is replaced by a constant is much more accurate for accreted crusts than for catalyzed crusts.

\section{Neutron-star properties}
\label{sec:ns-prop}

\subsection{Global structure of neutron stars}

The structure of a non-rotating NS is determined by the 
well-known Tolman-Oppenheimer-Volkoff (TOV) equations~\citep{tol39,op39}
\begin{eqnarray}
\label{eq:TOV1}
\frac{{\rm d}P(r)}{{\rm d}r} &=& -\frac{G\rho(r)\mathcal{M}(r)}{r^2}
\biggl[1+\frac{P(r)}{c^2\rho(r)}\biggr] \nonumber \\ 
&&\times \biggl[1+\frac{4\pi P(r)r^3}{c^2\mathcal{M}(r)}\biggr]\biggl[1-\frac{2G\mathcal{M}(r)}{c^2 r}\biggr]^{-1}
\end{eqnarray}
and
\begin{equation}
\label{eq:TOV2}
\mathcal{M}(r) = 4\pi\int_0^r\rho(r')r'^2{\rm d}r'\quad .
\end{equation}

The numerical solution of the TOV Eqs.~(\ref{eq:TOV1}) and (\ref{eq:TOV2}) is found by integrating from the centre of the star at $r=0$ up to the surface at (circumferential) radius $r=R$ such that $P(R)=0$. The gravitational mass of the star is then given by $M \equiv\mathcal{M}(R)$. In practice, the TOV equations were integrated down to the density $\rho=10^6$~g~cm$^{-3}$.

\begin{figure}
\begin{center}
\resizebox{\hsize}{!}{\includegraphics{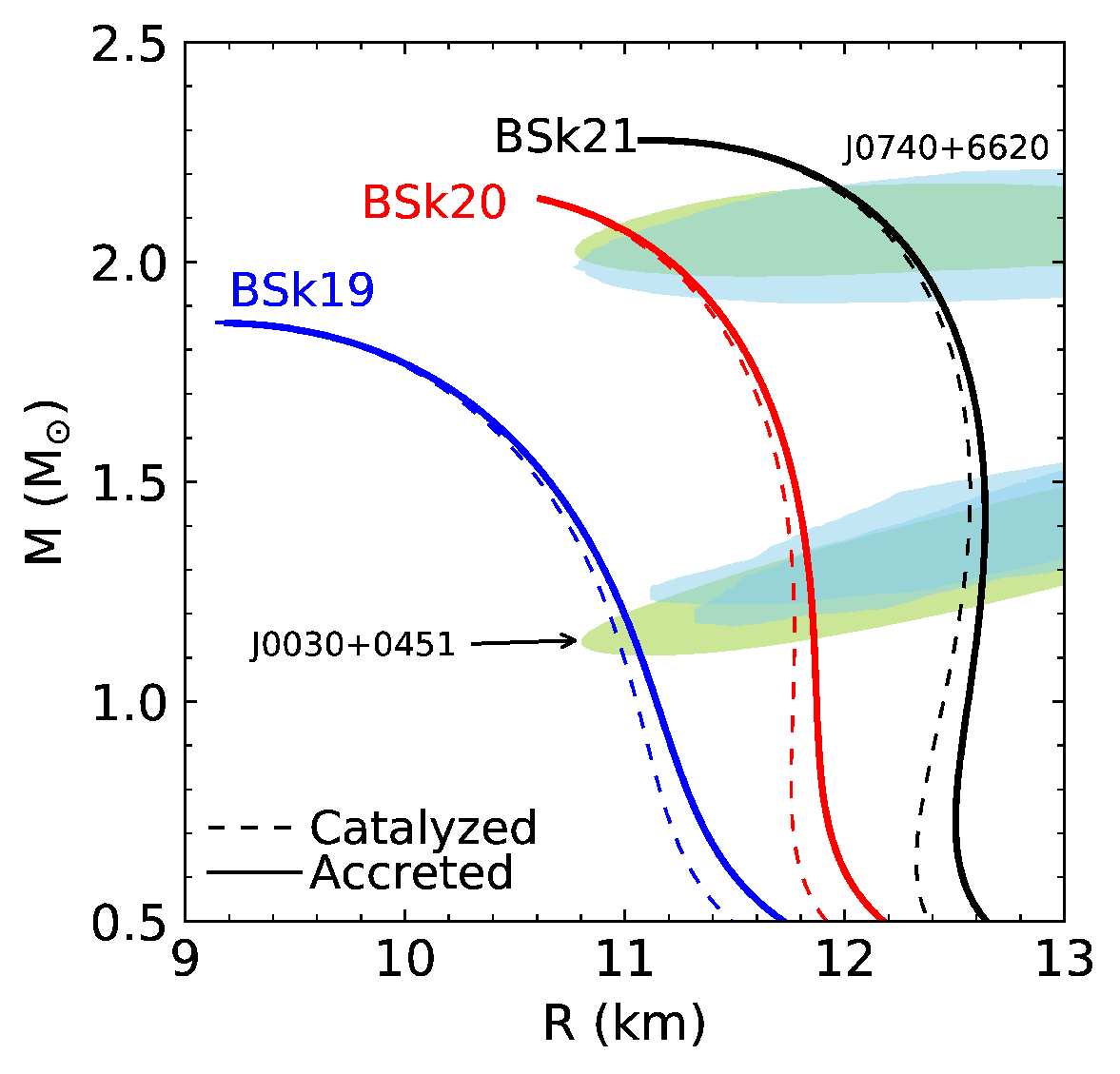}}
\end{center}
\caption{Mass-radius relation for accreted and catalyzed NSs for the EoSs based on the BSk19, BSk20, and BSk21 EDFs. Results obtained from Eq.~\eqref{eq:deltaR} are indistinguishable from the exact ones. The coloured areas represent the recent mass-radius measurements (at 1$\sigma$) from NICER of PSR~J0030$+$0451 and PSR~J0740$+$6620. The analyses of \cite{Riley2019, Riley2021} are shown in green and those of \cite{Miller2019, Miller2021} in blue. See text for details.
}
\label{fig:mrap}
\end{figure}

Numerical results are shown in Fig.~\ref{fig:mrap} for both accreted and catalyzed NSs for the EoSs based on the EDFs BSk19, BSk20, and BSk21.
We checked that the approximation proposed by \citet{zdunik2017} for determining the mass-radius relation from the core EoS only remains equally accurate for the EoSs calculated here.
As can be seen in Fig.~\ref{fig:mrap}, accreted NSs (solid lines) have larger radii than their non-accreted relatives made of catalyzed matter (dashed lines). This stems from the fact that the EoS of accreted NSs is stiffer, as discussed in Sect.~\ref{sec:EOS}. 
The differences in the radii is approximately given by \citep{zdunik2017}
\begin{equation}
\delta R\approx 144\ \left(\frac{Q_{\rm tot}}{2\ \mathrm{MeV}}\right) \left( \frac{R}{10\ \mathrm{km}} \right)^2\ 
  \left( \frac{M_\odot}{M} \right)\ 
  \left( 1 - \frac{2GM}{R c^2} \right)\ \mathrm{m} \ ,
  \label{eq:deltaR}
\end{equation}
where $R$ is the radius of the catalyzed NS and $Q_{\rm tot}$ is the total energy per nucleon released in accreted crust. 

For comparison, we show in Fig.~\ref{fig:mrap} the recent mass and radius estimates from NICER observations of PSR~J0030$+$0451  \citep{Miller2019,Riley2019} and PSR~J0740$+$6620  \citep{Miller2021, Riley2021}. Contours for the source PSR~J0740$+$6620 include XMM-Newton priors on the mass. The contours taken from \cite{Miller2019} include both the two and three oval-spot analyses of the signal. 
PSR~J0030$+$0451 and PSR~J0740$+$6620 are two millisecond pulsars and are therefore expected to have been recycled by accretion from a companion in their past evolution. The EoS of accreted NSs should thus be more realistic than that of catalyzed matter. However, the deviations $\delta R$ 
remain much smaller than the current observational uncertainties on the radii. 
Still, these observations clearly favour the EoSs based on the BSk20 and BSk21 EDFs. Incidentally, the latter is the one that was previously found to be in best agreement with nuclear data~\citep{goriely2010} and various other  astrophysical observations~\citep{fantina2013}. 
Even though the EoS based on the BSk19 EDF fails to reproduce the mass and radius of the very massive pulsar PSR~J0740$+$6620, it remains compatible with those of the less massive pulsar PSR~J0030$+$0451. This comparison is consistent with the analysis of GW170817.

\subsection{Moment of inertia}
\label{sect:Icrust}

The moments of inertia $I$ of accreted and catalyzed NSs were calculated in the slow rigid rotation approximation~\citep{hartle1967}, as implemented in \cite{haensel1982}\footnote{We note that in the second-to-last expression of Eq.~(13) in \citet{haensel1982} a factor 6 is missing.}: 
\begin{eqnarray}\label{eq:moment-inertia}
I&=&\frac{8\pi}{3}\int_0^R{\rm d}r\, r^4\biggl[\rho(r)+\frac{P(r)}{c^2}\biggr]e^{-\Phi(r)}\nonumber \\
&&\times \left(\sqrt{1-\frac{2G\mathcal{M}(r)}{c^2 r} }\right)^{-1} \frac{\bar \omega(r)}{\Omega}\quad , 
\end{eqnarray}
\begin{equation}
\label{eq:slow-rotation-j1}
\frac{{\rm d}\bar\omega (r)}{{\rm d}r}=\frac{6 G e^{\Phi(r)}}{c^2 r^4} \left(\sqrt{1-\frac{2G\mathcal{M}(r)}{c^2 r} }\right) ^{-1}j(r) \quad ,
\end{equation}
\begin{equation}
\label{eq:slow-rotation-j2}
\frac{{\rm d}j(r)}{{\rm d}r} = \frac{8\pi }{3} r^4\biggl[\rho(r)+\frac{P(r)}{c^2}\biggr]e^{-\Phi(r)}\left(\sqrt{1-\frac{2G\mathcal{M}(r)}{c^2 r} }\right)^{-1}  \bar\omega \quad ,
\end{equation}
where $\bar\omega(r)$ is the local spin frequency as measured in a local inertial frame, $j(r)$ represents the contribution from the sphere of radius $r$ to the stellar angular momentum $J=j(R)$, $\Omega$ is the (uniform) angular frequency of the star expressible as 
\begin{equation}
\Omega=\bar\omega(R)+\frac{2GJ}{c^2R^3}\quad ,
\end{equation}
and we have introduced the metric function $\Phi(r)$ 
defined by 
\begin{equation} 
\frac{{\rm d}\Phi}{{\rm d}r}=-\frac{1}{\rho(r) c^2}\frac{{\rm d}P}{{\rm d}r}\biggl[1+\frac{P(r)}{\rho(r) c^2}\biggr]^{-1}\quad .
\end{equation} 

\begin{figure}
\begin{center}
\resizebox{\hsize}{!}{\includegraphics{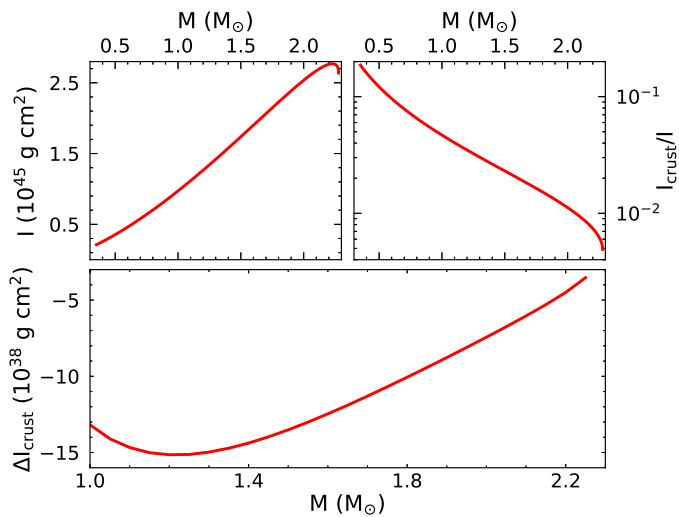}}
\end{center}
\caption{Moment of inertia of accreted NSs. Top panels: Total moment of inertia $I$ (left panel) and fractional moment of inertia of the crust $I_{\rm crust}/I$ (right panel) in cgs units as a function of the NS mass. Results with accreted or catalyzed crust are indistiguishable. Bottom panel: difference $\Delta I_{\rm crust}$ of the crustal moment of inertia between accreted and catalyzed NSs. Calculations were performed for the EoS based on the BSk21 EDF.} 
\label{fig:im}
\end{figure}

Results for the BSk21 EDF are shown in the top left panel of Fig.~\ref{fig:im} for different NS masses. The differences $\Delta I$ in the moment of inertia between accreted and catalyzed NSs are indistinguishable and are of the order of $\Delta I~\sim 10^{-7} I$. 

Of particular interest for the analysis of pulsar frequency glitches~\citep{andersson2012,chamel2013} is the fractional moment of inertia of the crust $I_{\rm crust}/I$, with 
\begin{eqnarray}\label{eq:crustal-moment-inertia}
I_{\rm crust}&=&\frac{8\pi}{3}\int_{r_{\rm cc}}^R{\rm d}r\, r^4\biggl[\rho(r)+\frac{P(r)}{c^2}\biggr]e^{-\Phi(r)}\nonumber \\
&&\times \left(\sqrt{1-\frac{2G\mathcal{M}(r)}{c^2 r} }\right)^{-1} \frac{\bar \omega(r)}{\Omega}\quad , 
\end{eqnarray}
$r_{\rm cc}$ denoting the radial coordinate at the crust-core boundary determined by the pressure $P_{\rm cc}=4.294\times 10^{32}$~dyn~cm$^{-2}$ \citep{pearson2012}. Results are shown in the top right panel of Fig.~\ref{fig:im}. Whether the crust is accreted or not hardly changes the crustal moment of inertia. The deviations $\Delta I_{\rm crust}$ plotted in the bottom panel of Fig.~\ref{fig:im} amount to $0.1\%$ of $I_{\rm crust}$ at most.

\subsection{Tidal deformability}
\label{sect:tidal}

\begin{figure}
\begin{center}
\resizebox{\hsize}{!}{\includegraphics{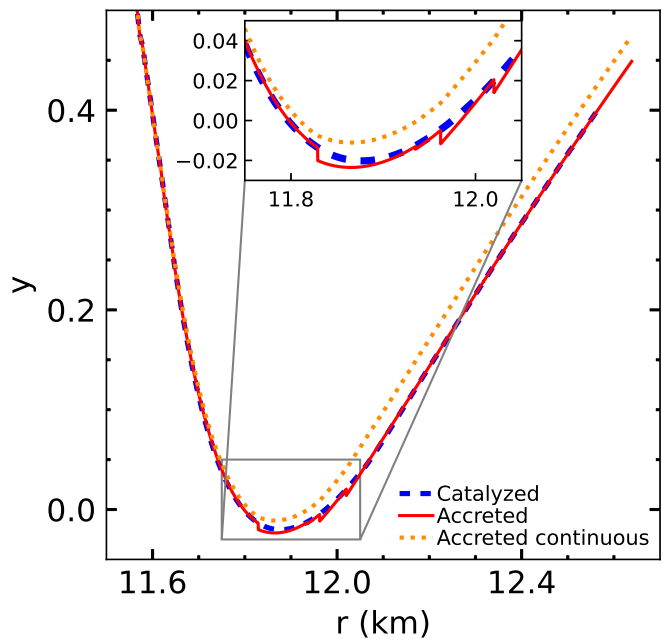}}
\end{center}
\caption{Function $y(r)$ characterising the tidal deformability of a $1.4\ M_\odot$ NS for values of the radial coordinate $r$ corresponding to the outer region of the star. For accreted NS, two approaches are presented: 1) treating $y(r)$ as a continuous function (dotted line) and 2) taking into account discontinuities (solid line). Results are compared to those obtained for catalyzed NS (dashed line). Calculations were performed for the EoS based on the BSk21 EDF.}
\label{fig:yr}
\end{figure}

\begin{figure}
\begin{center}
\resizebox{\hsize}{!}{\includegraphics{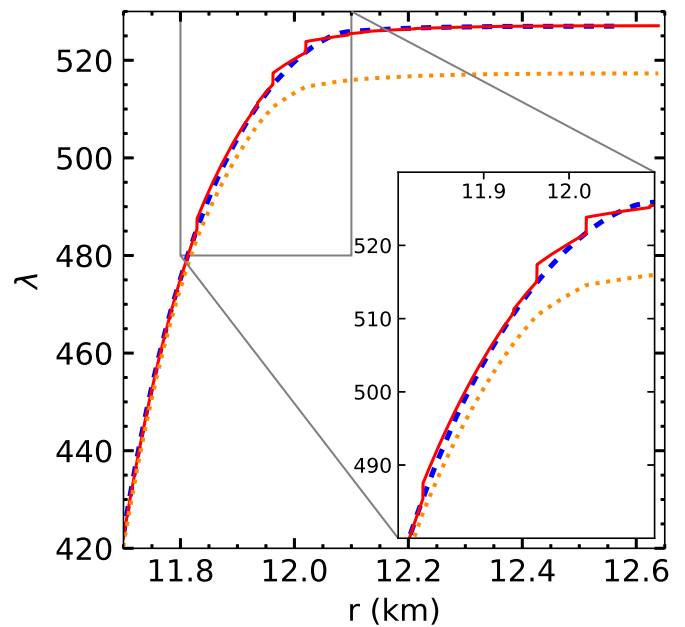}}
\end{center}
\caption{Same as Fig.~\ref{fig:yr} for the tidal deformability function $\lambda(r)$. 
}
\label{fig:lamr}
\end{figure}

To assess the importance of accretion on the tidal deformability of a NS, we calculated the tidal deformability function:
\begin{equation}
  \lambda(r) = \frac{2}{3} k_2(r) \left[\frac{rc^2}{G\mathcal{M}(r)}\right]^5\, , 
\label{eq:lambda} 
\end{equation} 
with the tidal Love number function $k_2(r)$ given by~\citep{hinderer2008,hinderer2010}
\begin{align}
k_2 =& \,\frac{8C^5}{5}(1-2C)^2\big[2+2C(y-1)-y\big]\nonumber \\
&\times \Big\{2C\big[6-3y+3C(5y-8)\big] \nonumber \\ 
&+ 4C^3\big[13-11y+C(3y-2)+2C^2(1+y)\big] \nonumber \\ 
&+ 3(1-2C)^2\big[2-y+2C(y-1)\big] \ln (1-2C) \Big\}^{-1} \, ,
\label{eq:k2}
\end{align}
where we have introduced the dimensionless compactness parameter
\begin{equation}
    C(r) = \dfrac{G\, \mathcal{M}(r)}{r\, c^2}\, ,
\end{equation}
and the function $y(r)$ is obtained by integrating the following differential equation with the boundary condition $y(0)=2$~\citep{postnikov2010}: 
\begin{equation}
\label{y_eq}
r \dfrac{ {\rm d} y}{{\rm d} r}+ y(r)^2 + F(r) y(r) + Q(r)=0 \, ,
\end{equation}
\begin{equation}
\label{F(r)}
    F(r)=\frac{1-4\pi G r^2(\mathcal{E}(r)-P(r))/ c^4}{1-2G\mathcal{M}(r)/(r c^2)}\, ,
\end{equation}
\begin{align}
\label{Q(r)}
    Q(r)=&\frac{4 \pi G r^2/c^4}{1-2G\mathcal{M}(r)/(r c^2)}\Biggl[5\mathcal{E}(r)+9P(r)\nonumber \\
    &+\frac{\mathcal{E}(r)+P(r)}{c_s(r)^2} c^2-\frac{6\,  c^4}{4\pi r^2 G}\Biggr] \nonumber \\
    &-4\Biggl[ \frac{G(\mathcal{M}(r)/(r c^2)+4\pi r^2 P(r)/c^4)}{1-2G\mathcal{M}(r)/(r c^2)}\Biggr]^2\, ,
\end{align}
$c_s=c\sqrt{ {\rm d} P/ {\rm d} \mathcal{E}}$ denoting the sound speed. Results for the functions $y(r)$ and $\lambda(r)$ for a NS with a gravitational mass $M=1.4\ M_\odot$ are plotted in Figs.~\ref{fig:yr} and \ref{fig:lamr} respectively, focussing on the outer region where the EoS of accreted NSs differs from that of catalyzed NSs. Calculations were made for the EoS based on the BSk21 EDF. Accreted NSs are found to have a slightly smaller tidal deformability than catalyzed NSs. However, the deviations almost completely disappear when density discontinuities at the interfaces between adjacent crustal layers are properly taken into account following the prescription of \cite{damour2009}, as can be seen by comparing the solid lines and the dashed lines.

\begin{figure}
\begin{center}
\resizebox{\hsize}{!}{\includegraphics{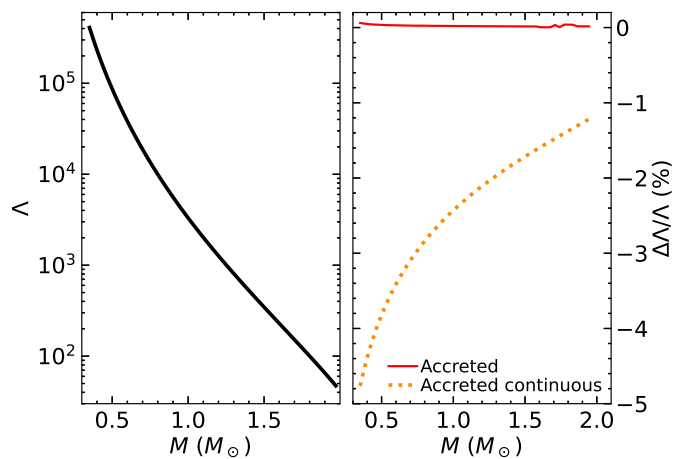}}
\end{center}
\caption{Tidal deformability of accreted NSs. Left panel: Tidal deformability $\Lambda$ as a function of the NS mass. Right panel: Relative deviation between the value of $\Lambda$ for accreted and catalyzed NSs with (solid line) and without (dotted line) a proper treatment of density discontinuities. Calculations were performed for the EoS based on the BSk21 EDF.}
\label{fig:deltalm}
\end{figure}

The (observable) tidal deformability of the star given by $\Lambda=\lambda(R)$ is displayed in the left panel of Fig.~\ref{fig:deltalm} for different NS masses using the EoS based on the BSk21 EDF.
Results for accreted and catalyzed NSs are indistinguishable. The relative deviations are shown in the right panel. They are of the same order $10^{-5}-10^{-6}$ as the precision of the calculations with the prescription of \cite{damour2009}. 
The Love numbers and tidal deformability coefficients previously calculated for catalyzed matter with the EoSs based on the BSk19, BSk20, and BSk21 EDFs as well as the comparison with the LIGO/Virgo measurement of GW170817 \citep{perot2019} therefore remain valid for accreted NSs. Ignoring density discontinuities leads to errors of the order of a few \% (below 2\% for NSs with masses $M\gtrsim M_\odot$); although these errors remain rather small, they are larger than the relative contribution of the crust to $\Lambda$ \citep{perot2020}. A proper treatment of discontinuities is therefore necessary to correctly assess the role of the crust.

\section{Conclusions}
\label{sect:discussion}

In this paper, we presented the EoSs for accreted NSs, calculated employing the Brussels-Montreal EDFs BSk19, BSk20, and BSk21, and we evaluated their impact on global NS properties.
Both the outer and inner crusts and the core were described with the same functional, thus ensuring a unified and thermodynamically consistent treatment.
The EoS of accreted NS crusts is found to be remarkably different from that of catalyzed crusts, the former being notably stiffer than the latter, particularly in the density region between the neutron drip and $\approx 10^{13}$~g~cm$^{-3}$.
This reflects in the adiabatic index, whose value $\approx 4/3$ remains rather constant throughout the accreted crust, implying that the polytropic approximation is much more accurate for accreted crusts than for catalyzed ones.

The different stiffness between accreted and catalyzed EoSs also impacts the global NS structure.
Indeed, accreted NSs with low and intermediate masses have larger radii than the non-accreted counterparts.
However, the deviations 
remain smaller than current observational uncertainties on the radii. Likewise,
the crustal moment of inertia and the tidal deformability of accreted NSs are almost indistinguishable from those of non-accreted NSs, provided density discontinuities at the interface between adjacent crustal layers are properly taken into account.

Although we considered here three specific models, to a good approximation the EoSs calculated for catalyzed matter remain applicable to compute the global properties of NSs, since they are mainly determined by the core. However, the composition of accreted crust, the heat sources, and their location may change depending on the chosen nuclear model. The BSk19, BSk20, and BSk21 EDFs were fitted to different realistic neutron-matter EoSs but were all constrained to yield the same symmetry energy coefficient $J$. 
The influence of varying this nuclear-matter parameter in accreted NSs has only been investigated on the boundary between the outer and inner crust in \citet{fantina2016}. The role of the symmetry energy in the crustal heating is left for future studies.

The results presented in this work complement our calculations of unified EoSs of catalyzed matter \citep{pearson2011, pearson2012, fantina2013, potekhin2013}, of tidal deformability of NSs \citep{perot2019,perot2020,perot2021}, and of the crustal heating in accreted NSs \citep{fantina2018}, thus providing consistent microscopic inputs based on the same EDFs for simulations of both accreting and non-accreted NSs. 
In our calculations, we did not consider neutron diffusion in the inner crust of accreting NSs \citep{shch2018, chu2020, gus2020}. Its inclusion within the microscopic approach adopted here is left for future studies.

\begin{acknowledgements}
This work was partially supported by the Polish National Science Centre (NCN)  grant no. 2018/29/B/ST9/02013, the CNRS PICS07889, the CNRS International Research Project (IRP) ``Origine des \'el\'ements lourds dans l'univers: Astres Compacts et Nucl\'eosynth\`ese (ACNu)'', the Fonds de la Recherche Scientifique (Belgium) under Grant No. IISN 4.4502.19, and the European Cooperation in Science and Technology (COST) action CA16214. S.G. and N.C. are F.R.S.-FNRS senior research associates. 
\end{acknowledgements}


\end{document}